\documentclass[preprint,showpacs,preprintnumbers,amsmath,amssymb,pra]{revtex4}

\usepackage{epsf}
\usepackage{dcolumn}
\usepackage{bm}
\everymath{\displaystyle}
\begin{document}

\title{General method of the relativistic Foldy-Wouthuysen transformation and proof of validity of the Foldy-Wouthuysen Hamiltonian}

\author{Alexander J. Silenko}

\affiliation{Research Institute for Nuclear Problems, Belarusian
State University, 220030 Minsk, Belarus\\
Bogoliubov Laboratory of Theoretical Physics, Joint Institute for
Nuclear Research, Dubna 141980, Russia}

\date{\today}

\begin {abstract}
A general method of the Foldy-Wouthuysen transformation is developed. This method is applicable to relativistic particles with any spin in arbitrarily strong external fields. It can be used when the de Broglie wavelength is much smaller than the characteristic distance.
Contrary to previously developed relativistic methods, the present method satisfies the condition of the exact Foldy-Wouthuysen transformation and is well substantiated. The derived relativistic Foldy-Wouthuysen Hamiltonian is expanded in powers of the Planck constant. In this expansion, terms proportional to the zero and first
powers are determined exactly in accordance with the above condition and terms proportional to higher
powers are not specified. The obtained result agrees with the corresponding formula for the Foldy-Wouthuysen Hamiltonian previously deduced by an iterative relativistic method and proves the validity of results obtained with this formula.
\end{abstract}

 \maketitle

\section{Introduction}

The Foldy-Wouthuysen (FW) representation \cite{FW} occupies
a special place in quantum theory due to its unique properties. In this representation, the Hamiltonian and all operators are
block-diagonal (diagonal in two spinors). Relations between
the operators in the FW representation are similar to those
between the respective classical quantities. Quantum-mechanical operators for relativistic particles
in external fields have the same form as those in the nonrelativistic quantum theory. In particular, the position
operator \cite{NW} and the momentum one are equal to $\bm r$ and $\bm p=-i\hbar\nabla$. The polarization operator for
spin-1/2 particles is defined by the Dirac matrix $\bm \Pi$. In other representations, these operators are expressed by much more cumbersome
formulas (see \cite{FW,JMP}). A great advantage of the FW representation is the simple
form of operators corresponding to classical observables. Thanks to these
properties, the FW representation provides
the best possibility of obtaining a meaningful classical limit
of the relativistic quantum mechanics \cite{FW,CMcK}.
The passage to the classical limit usually reduces to a replacement of the operators in quantum-mechanical Hamiltonians and equations of motion
with the corresponding classical quantities. The possibility of such a replacement, explicitly
or implicitly used in practically all works devoted to the relativistic FW transformation, was recently
rigorously proved in Ref. \cite{JINRLett12}.

Since the FW representation is very important for contemporary quantum mechanics and elementary particle
physics, the problem of passing to this representation is very actual. A diagonalization of a Hamiltonian does not necessarily lead to the FW representation \cite{JMPcond}. The condition uniquely defining the FW representation has been formulated by Eriksen \cite{E}.
There is an infinite set of representations different from the FW representation while their distinctive feature is a block-diagonal form of the Hamiltonian. Paradoxically,
even the original method by Foldy and Wouthuysen \cite{FW} is approximate and does not lead to the FW representation \cite{EK,dVFor}.
The Eriksen method \cite{E} is exact but it explicitly represents the FW Hamiltonian as a series of relativistic corrections to a nonrelativistic limit and does not give a compact
relativistic expression for this Hamiltonian. Nevertheless, the series of relativistic corrections as a whole defines the \emph{exact} FW Hamiltonian. Some iterative (``step-by-step'') methods of the FW transformation widely used in
quantum chemistry \cite{QuantChem} possess similar properties. They also express the exact FW Hamiltonian as a series of relativistic corrections satisfying the Eriksen condition.

Compact relativistic expressions for the FW Hamiltonian are necessary to describe high-energy particles (e.g., in accelerators and storage rings). In this case, the series of relativistic corrections is insufficient, and a compact relativistic formula is needed. Moreover, this series becomes divergent on the condition that $\bm p^2/(mc)^2\geq1$. In such cases, the FW transformation can be carried out by relativistic methods. However, the known relativistic methods of the FW transformation \cite{B,JMP,TMP,BliokhK,PRA,Goss} are approximate. A determination of their precision and of limits of their applicability has been made in Ref. \cite{TMPFW}. Unfortunately, these methods are not well substantiated. In particular, their accordance with the Eriksen condition was never investigated.

In the present work, we propose a new method of the FW transformation. The method is general because it is applicable to relativistic particles with any spin in arbitrarily strong external fields. Contrary to previously developed relativistic methods, it satisfies the condition of the exact FW transformation.
We rigorously derive the general formula for the relativistic FW Hamiltonian which is expanded in powers of the Planck constant. In this expansion, terms proportional to the zero and first
powers are determined exactly in accordance with the above condition. The obtained result agrees with the corresponding formula for the FW Hamiltonian previously deduced by the iterative relativistic method \cite{PRA} and therefore proves the validity of results obtained with this formula. Like other relativistic methods applying an expansion of the FW Hamiltonian in powers of the Planck constant \cite{BliokhK,PRA,Goss}, the proposed method can be used when the de Broglie wavelength is much smaller than the characteristic distance.


\section{Exact and approximate methods of the Foldy-Wouthuysen transformation}\label{Methods}

We extend the Eriksen method of the exact FW transformation \cite{E} to particles with any spin $s$. As a rule, the corresponding wave functions can be separated into two spinor-like blocks. The FW transformation brings the initial Hamiltonian to a block-diagonal form. One of the blocks of the FW wave function (upper and
lower blocks for states with positive and negative total energy, respectively) is equal to zero.
The operator transforming the initial Hamiltonian to the FW representation can be presented in the exponential form:
\begin{equation} U_{FW}=\exp{(i{S})}.\label{Vvetott} \end{equation}
The even (block-diagonal) form of the final Hamiltonian was the only condition of transformation
used by Foldy and Wouthuysen \cite{FW}. However, this condition does
not define the FW Hamiltonian unambiguously. The additional condition eliminating this ambiguity has been proposed by Eriksen \cite{E} and substantiated by Eriksen and Kolsrud \cite{EK}. Additional substantiation of the Eriksen method has been given in Ref. \cite{VJ}. The transformation remains unique if the
operator ${S}$ in Eq. (\ref{Vvetott}) is \emph{i)} odd,
\begin{equation} \beta{S}=-{S}\beta,
\label{VveEfrt} \end{equation} and \emph{ii)} Hermitian for fermions and
$\beta$-pseudo-Hermitian (${S}={S}^\ddag\equiv\beta{S}^\dag\beta$) for bosons. Here $\beta$ is the Dirac matrix for spin-1/2 particles. It takes the form $\beta=\rho_3\otimes\mathfrak{I}$ for particles with a different spin ($\rho_3$ is the Pauli matrix whose components act on spinor-like blocks of the wave function and $\mathfrak{I}$ is the $(2s+1)\times(2s+1)$ unit matrix).

Condition (\ref{VveEfrt}) is equivalent to \cite{E,EK}
\begin{equation} \beta U_{FW}=U^\dag_{FW}\beta.\label{Erikcon} \end{equation}
Thus, the FW transformation operator should satisfy Eq.
(\ref{Erikcon}) and should perform the transformation in one step.
Eriksen \cite{E} has found the operator possessing the needed
properties:
\begin{equation}
U_{E}=U_{FW}=\frac{1+\beta\lambda}{\sqrt{2+\beta\lambda+\lambda\beta}},
~~~ \lambda=\frac{{\cal H}}{({\cal H}^2)^{1/2}}, \label{eqXXI}
\end{equation}
where $\lambda$ is the sign operator. The two operator factors in the denominator commute. The denominator is an even operator and commutes with the numerator (see Refs. \cite{E,JMPcond,TMPFW}). While Eriksen has considered only a Dirac particle in stationary fields, Eq. (\ref{eqXXI}) remains valid for an arbitrary-spin particle in nonstationary fields. This statement is based on the fact that the operators $1+\beta\lambda$ and $U_{E}$ vanish either a lower or a upper spinor (or corresponding spinor-like blocks) for positive and negative energy states, respectively. The operator $U_{E}$ satisfies the Eriksen condition.
The general form of the Hamiltonian applicable for an arbitrary-spin particle is given by
\begin{equation} {\cal H}=\beta{\cal M}+{\cal E}+{\cal
O},~~~\beta{\cal M}={\cal M}\beta, ~~~\beta{\cal E}={\cal E}\beta, ~~~\beta{\cal O}=-{\cal
O}\beta, \label{eq3} \end{equation} where $\beta{\cal M}+{\cal E}$ and ${\cal O}$ are even and odd operators commuting and anticommuting with the operator
$\beta$, respectively. Similarly to the block structure of $\beta$, the block structure of any operator in Eq. (\ref{eq3}) can be presented in terms of the three Pauli matrices, $\rho_1,\rho_2,\rho_3$, and of the $2\times2$ unit matrix $\mathcal{I}$ multiplied by some $(2s+1)\times(2s+1)$ operators.

We may suppose that the operators $\beta{\cal M}$ and ${\cal E}$ can be presented in the form $\beta{\cal M}=\rho_3\otimes\mathfrak{M},~{\cal E}=\mathcal{I}\otimes\mathfrak{E}$. However, such a separation of the even part of ${\cal H}$ is not obligatory. We need to require only that the commutators
$[{\cal O},{\cal M}]$ and $[{\cal O},{\cal E}]$ have an additional factor $\hbar$ as compared with the corresponding products of operators, ${\cal O}{\cal M}$ and ${\cal O}{\cal E}$.

For electromagnetic interactions of a Dirac particle, the operator ${\cal M}$ is replaced with the particle mass:
\begin{equation} {\cal H}_D=\beta m+{\cal E}+{\cal
O}. \label{eq3Dirac} \end{equation}
In this case, the FW Hamiltonian can be represented as a series of
relativistic corrections in powers of the operators ${\cal E}/m$ and ${\cal
O}/m$. The Eriksen method cannot be used for obtaining compact
relativistic formulas for this Hamiltonian (except for some special cases \cite{JINREriksn}) because the general formula (\ref{eqXXI}) is very cumbersome
and contains square roots of Dirac matrices.

The needed series of relativistic corrections can be obtained analytically
with a computer \cite{VJ}. We present below the exact FW Hamiltonian
calculated by de Vries and Jonker up to terms of the order of
$(v/c)^8$. They considered the stationary case and supposed that ${\cal O}/m\sim v/c$ and ${\cal
E}/m\sim(v/c)^2$. The result of the calculations \cite{VJ} has been
presented in a more convenient form \cite{TMPFW} via multiple
commutators:
\begin{equation}\begin{array}{c}
{\cal H}_{FW}=\beta\left(m+ \frac {{\cal
O}^2}{2m}-\frac{{\cal O}^4}{8m^3}+\frac{{\cal
O}^6}{16m^5}-\frac{5{\cal O}^8}{128m^7}\right)\\+{\cal
E}-\frac{1}{128m^6}\left\{(8m^4-6m^2{\cal O}^2+5{\cal O}^4),[{\cal
O},[{\cal O},{\cal E}]]\right\}\\+\frac{1}{512m^6}\left\{(2m^2-{\cal O}^2),[{\cal
O}^2,[{\cal O}^2,{\cal E}]]\right\}\\
+\frac{1}{16m^3}\beta\left\{{\cal O},\left[[{\cal O},{\cal E}],{\cal E}\right]\right\}-\frac{1}{32m^4}\left[{\cal
O},\left[[[{\cal O},{\cal E}],{\cal E}],{\cal E}\right]\right]\\+\frac{11}{1024m^6}\left[{\cal O}^2,\left[{\cal
O}^2,[{\cal O},[{\cal O},{\cal E}]]\right]\right]+A_{24},
\end{array} \label{eq12erk}
\end{equation}
where
\begin{equation}\begin{array}{c} A_{24}=\frac{1}{256m^5}\beta\Biggl(24\left\{{\cal O}^2, ([{\cal O},{\cal E}])^2\right\}-
20\left ([{\cal O}^2,{\cal E}]\right )^2-14\left\{{\cal O}^2,
\left [[{\cal O}^2,{\cal E}],{\cal E}\right]
\right\}\\-4\left[{\cal O},\left[{\cal O},\left[[{\cal O}^2,{\cal E}],{\cal E}\right] \right] \right]+\frac92 \left[\left[{\cal
O},\left[{\cal O},\left[{\cal O}^2,{\cal E}\right]\right]
\right],{\cal E}\right]\\-\frac92 \left[\left[{\cal O},\left[{\cal
O},{\cal E}\right]\right],\left[{\cal O}^2,{\cal E}\right]\right]
+\frac52 \left [{\cal O}^2,\left[{\cal O},\left[[{\cal O},{\cal E}],{\cal E}\right] \right] \right]\Biggr).
\end{array} \label{eq12A}
\end{equation}
In $A_{24}$, the first and second subscripts indicate the
respective numbers of operators ${\cal E}$ and ${\cal O}$ in the
product. A mistake in the calculation of this term made in Ref.
\cite{TMPFW} is corrected here.

Terms of higher orders up to $(v/c)^{12}$ have also been
calculated many years ago (see Ref. \cite{VJ} and references
therein). This fact clearly demonstrates the applicability and
importance of the Eriksen method for a calculation of the FW
Hamiltonian as a series of relativistic corrections.

The Eriksen condition is general and is applicable to an arbitrary-spin particle. However, this condition was used neither by Eriksen nor in other works even for a relativistic/ultrarelativistic Dirac particle, let alone for a particle with another spin.

The problem of description of single relativistic and ultrarelativistic
particles in external fields may be solved
with relativistic methods of the FW transformation. These methods use either the weak field approximation \cite{B,JMP,TMP} or an expansion of the FW Hamiltonian
into a power series in the Planck constant \cite{BliokhK,PRA,Goss}. All known relativistic methods are iterative and approximate. A determination of limits of their applicability \cite{TMPFW} has been based on a comparison of the FW Hamiltonians obtained with the relativistic methods and with the Eriksen method. However, such a
comparison can be fulfilled only for Dirac particles because the particle mass cannot be replaced by the operator ${\cal M}$ in an expansion in powers of the operators ${\cal E}/m$ and ${\cal O}/m$.

To fulfill the FW transformation of the initial Hamiltonians (\ref{eq3}) and (\ref{eq3Dirac}), a priori information about commutation relations should be used. Any commutator of the momentum and coordinate operators adds the factor $\hbar$, while a commutator of different Pauli (or Dirac) matrices does not affix such a factor. Thus, we suppose that commutators like
$[{\cal O},{\cal M}]$ and $[{\cal O},{\cal E}]$ have the additional factor $\hbar$ as compared with the products of the operators, ${\cal O}{\cal M}$ and ${\cal O}{\cal E}$. Odd terms contain off-diagonal Pauli matrices, $\rho_1$ and $\rho_2$, acting on spinors (or spinor-like blocks of bispinor-like wave functions for particles with other spins). Since these matrices do not commute with each other, we assume that
multiple commutators of the form $[{\cal O},[{\cal O},\dots[{\cal O},{\cal E}]\dots]]$ are of the order of $\hbar/S_0$ with respect to the operator
product ${\cal O}{\cal O}\dots{\cal O}{\cal E}$ with the factor $\hbar$ already appearing due to the first commutation. Here $S_0$ is some quantity with the
dimension of action (see Ref. \cite{TMPFW} for details). In contrast, commutators of the forms $[{\cal O}^2,[{\cal
O},{\cal E}]]$, $[{\cal O}^2,[{\cal O}^2,{\cal E}]]$, and $[[{\cal
O},{\cal E}],{\cal E}]$ are of the order of $(\hbar/S_0)^2$ (with respect to the product of the operators appearing
in them). This property takes place because ${\cal
O}^2$ and ${\cal E}$ are even (block-diagonal) operators proportional to the $2\times2$ unit matrix. Of course, this a priori information should be verified in any specific case.

When determining the order of magnitude, we indicate the smallest possible degree in $\hbar$. For example, the commutator $[{\cal O},[{\cal O},{\cal E}]]$ can be of the order of $(\hbar/S_0)^2$ instead of $\hbar/S_0$. The
commutator $[{\cal O},{\cal E}]$ with the nominal order of $\hbar/S_0$ can additionally contain terms of the orders of
$(\hbar/S_0)^2, (\hbar/S_0)^3, \dots$, can be of the second or higher order in $\hbar/S_0$, or can be equal to zero.

The 
general relativistic method proposed in Ref. \cite{PRA} is applicable for an arbitrary-spin particle and transforms the Hamiltonian (\ref{eq3}) to the FW representation in the nonstationary case. The first iteration is
made with the transformation operator
\begin{equation}
U=\frac{\beta\epsilon+\beta {\cal M}-{\cal
O}}{\sqrt{(\beta\epsilon+\beta {\cal M}-{\cal O})^2}}\,\beta,~~~
U^{-1}=\beta\,\frac{\beta\epsilon+\beta{\cal M}-{\cal
O}}{\sqrt{(\beta\epsilon+\beta{\cal M}-{\cal O})^2}}.
\label{eq18N} \end{equation} Here $U^{-1}=U^\dagger$ when the Hamiltonian (\ref{eq3}) is Hermitian (${\cal
H}={\cal H}^\dagger$), $U^{-1}=U^\ddagger$ when it is $\beta$-pseudo-Hermitian (${\cal H}={\cal
H}^\ddagger$), and
\begin{equation}\epsilon=\sqrt{{\cal M}^2+{\cal
O}^2}.\label{epsln} \end{equation}

We consider the general case when external fields are
nonstationary. In this
case, the exact formula for the transformed Hamiltonian has the form
\begin{equation} \begin{array}{c} {\cal H}'=\beta\epsilon+{\cal
E}+ \frac{1}{2T}\Biggl(\left[T,\left[T,(\beta\epsilon+{\cal
F})\right]\right] +\beta\left[{\cal O},[{\cal O},{\cal
M}]\right]\\- \left[{\cal O},\left[{\cal O},{\cal F}\right]\right]
- \left[(\epsilon+{\cal M}),\left[(\epsilon+{\cal M}),{\cal
F}\right]\right] - \left[(\epsilon+{\cal M}),\left[{\cal M},{\cal
O}\right]\right]\\-\beta \left\{{\cal O},\left[(\epsilon+{\cal
M}),{\cal F}\right]\right\}+\beta \left\{(\epsilon+{\cal
M}),\left[{\cal O},{\cal F}\right]\right\} \Biggr)\frac{1}{T},
\end{array} \label{eq28N} \end{equation} where ${\cal F}={\cal E}-i\hbar\frac{\partial}{\partial
t}$ and
$T=\sqrt{(\beta\epsilon+\beta{\cal M}-{\cal O})^2}$.

Hamiltonian (\ref{eq28N}) still contains odd terms proportional to
the first and higher powers of the Planck constant. Additional transformations
bring it to the block-diagonal form. If one holds only terms proportional to the zero and first powers of $\hbar$, the final FW Hamiltonian takes the form
\begin{equation}\begin{array}{c}
{\cal H}_{FW}=\beta\epsilon+ {\cal E}+\frac 14\left\{\frac{1}
{2\epsilon^2+\{\epsilon,{\cal M}\}},\left(\beta\left[{\cal O},[{\cal O},{\cal
M}]\right]-[{\cal O},[{\cal O},{\cal
F}]]\right)\right\}. \end{array} \label{eqfingn}
\end{equation}
It should be emphasized that Eqs. (\ref{eq18N})--(\ref{eqfingn}) are applicable for arbitrarily
strong external fields. For a Dirac particle, ${\cal M}=m$.

However, this method was substantiated only indirectly, namely, by its agreement with the similar method developed for a Dirac particle and by the consistency of the obtained FW Hamiltonians with the corresponding classical Hamiltonians.

The related relativistic method of the FW transformation for a Dirac particle \cite{JMP} uses both the exponential and nonexponential transformation operators.
In Ref. \cite{TMPFW}, this method has been 
substantiated (for $\bm p^2/(mc)^2<1$) and the limits of its applicability have been determined. It has been shown that terms in the relativistic FW Hamiltonians proportional to the zero and first powers of $\hbar$ are exact within the precision of the analysis. In Ref. \cite{TMPFW}, relativistic corrections up to the order of $(v/c)^8$ (when ${\cal E}/m\sim(v/c)^2, ~ {\cal O}/m\sim v/c$) obtained with the Eriksen method have been taken into account.
The FW transformations of the Dirac and Dirac-Pauli Hamiltonians for a particle 
in electric and magnetic fields have been fulfilled in Ref. \cite{ChenChiou} with the Kutzelnigg diagonalization method \cite{Kutzelnigg} up to terms of $(|\bm\pi|/mc)^{14}\approx(v/c)^{14}$, where $\bm\pi$ is the kinetic momentum. This method satisfies the Eriksen condition and is applicable in the weak-field approximation. The FW Hamiltonians calculated in Ref. \cite{ChenChiou} are in agreement with the corresponding relativistic Hamiltonians derived in Refs. \cite{JMP,TMPFW} as well as with the classical ones. Thus, the important results obtained in Ref. \cite{ChenChiou} also confirm that the relativistic method \cite{JMP} gives FW Hamiltonians exact for terms proportional to the zero and first powers of $\hbar$.

The presented analysis nevertheless shows that the main disadvantage of all known relativistic methods \cite{B,JMP,TMP,BliokhK,PRA,Goss} is a lack of their
substantiations. The most important substantiation \cite{TMPFW} of the method developed for single Dirac particles \cite{JMP} is the full
agreement of the obtained relativistic FW Hamiltonian (\ref{eq12A}) with the exact FW Hamiltonian calculated with the Eriksen method and
expressed by Eqs. (\ref{eq12erk}) and (\ref{eq12A}) as a series of relativistic corrections. However, such a substantiation can be made only
on the condition that $\bm p^2/(mc)^2<1$. The consistency of FW Hamiltonians obtained with the method \cite{JMP} and with the other methods developed for single Dirac particles \cite{B,TMP,BliokhK,Goss} demonstrates the validity of the
latter on this condition
(for details, see Ref. \cite{TMPFW}).

The only relativistic method applicable to single arbitrary-spin particles \cite{PRA} is substantiated by the consistency of the FW transformation operator used with the corresponding operator for a Dirac particle \cite{JMP}. The validity of all relativistic methods is supported by an agreement between obtained quantum-mechanical Hamiltonians in the FW representation and corresponding classical Hamiltonians. But this substantiation is not satisfactory.

In the next sections, we will expound a new 
method of the relativistic FW transformation for single arbitrary-spin particles. In contrast to all known relativistic methods, this method is based on the fulfillment of the Eriksen condition of the exact FW transformation
presented in this section. The developed method exploits the exponential transformation operator and is general because it is applicable for arbitrarily strong interactions.

\section{Approximate Foldy-Wouthuysen transformation for a relativistic arbitrary-spin particle}

Let us choose the exponential transformation operator as follows:
\begin{equation}\begin{array}{c}
{S}=-i\beta\Theta. \end{array} \label{exptrop}
\end{equation}
In this case
\begin{equation}\begin{array}{c}
U=\cos{{S}}+i\sin{{S}}. \end{array} \label{exptroe}
\end{equation}
Power series expansions of the sine and cosine functions show that
\begin{equation}\begin{array}{c}
U=\cos{\Theta}+i(-i\beta)\sin{\Theta}=\cos{\Theta}+\beta\sin{\Theta}. \end{array} \label{exptrol}
\end{equation}

The transformation of the initial Hamiltonian (\ref{eq3}) to another representation is given by
\begin{equation} {\cal H}'=i\hbar\frac{\partial}{\partial
t}+U\left({\cal H}-i\hbar\frac{\partial}{\partial
t}\right)U^{-1}=i\hbar\frac{\partial}{\partial
t}+U\left(\beta{\cal M}+{\cal E}+{\cal
O}-i\hbar\frac{\partial}{\partial
t}\right)U^{-1}.
\label{taeq3}
\end{equation}

When $[{\cal O},{\cal
M}]=[{\cal O},{\cal F}]=[{\cal M},{\cal F}]=0$, the transformation (\ref{exptrol}), (\ref{taeq3}) is exact and leads to the FW representation (cf. Ref. \cite{JMP}):
$$\begin{array}{c}
{\cal H}_{FW}=(\cos{\Theta}+\beta\sin{\Theta})\left(\beta{\cal M}+{\cal F}+{\cal
O}\right)(\cos{\Theta}-\beta\sin{\Theta})+i\hbar\frac{\partial}{\partial t}\\=
(\beta {\cal M}+{\cal
O})\left(\cos\Theta-\beta\sin \Theta\right)^2+
{\cal E} =(\beta {\cal M}+{\cal O})\left(\cos {2\Theta}-\beta\sin {2\Theta}\right)+{\cal E}\\= \beta\left({\cal M}\cos
{2\Theta}+{\cal O}\sin {2\Theta}\right)+{\cal O}\cos {2\Theta}-{\cal M}
\sin {2\Theta}+{\cal E}.\end{array}$$

The transformed Hamiltonian is even when
\begin{equation}\begin{array}{c}
\tan {2\Theta}=\frac{{\cal O}}{{\cal M}}. \end{array} \label{exctndn}
\end{equation}

The same transformation operator can be used in the general case. However, a noncommutativity of operators requires additional steps of transformation. The needed generalization of Eq. (\ref{exctndn}) to noncommutative operators ${\cal O}$ and ${\cal M}$ is given by
\begin{equation}\begin{array}{c}
2\Theta=\arctan{X}, ~~~ X=\left\{\frac{1}{2{\cal M}},{\cal O}\right\}. \end{array} \label{exctnew}
\end{equation}
The operator $\Theta$ is odd. It is also Hermitian for spin-1/2 particles and other fermions and $\beta$-pseudo-Hermitian for bosons. It is convenient to utilize the operator relations
\begin{equation}\begin{array}{c}
\sin{\Theta}=\frac{\tan {2\Theta}}{\sqrt{2\sqrt{1+\tan^2{2\Theta}}\left(1+\sqrt{1+\tan^2{2\Theta}}\right)}}, \end{array} \label{oprel}
\end{equation}
\begin{equation}\begin{array}{c} \cos{\Theta}=\frac{1+\sqrt{1+\tan^2{2\Theta}}} {\sqrt{2\sqrt{1+\tan^2{2\Theta}}(1+\sqrt{1+\tan^2{2\Theta}})}}. \end{array} \label{opren}
\end{equation} Operator relations (\ref{oprel}) and (\ref{opren}) repeat the corresponding trigonometrical ones.
In the general case, $\tan{\Theta}=
\tan {2\Theta}/(1\pm\sqrt{1+\tan^2 {2\Theta}})$, but only the plus sign corresponds to the
FW transformation \cite{JMP}.

The nonexponential transformation operator can be presented as follows:
\begin{equation}\begin{array}{c}
U=\frac{1+\sqrt{1+X^2}+\beta X}{\sqrt{2\sqrt{1+X^2}\left(1+\sqrt{1+X^2}\right)}}. \end{array} \label{nexptop}
\end{equation}

When we take into account only terms proportional to
the zero and first powers of $\hbar$, we obtain the following relation:
$$\begin{array}{c} \left(1+\sqrt{1+X^2}+\beta X\right)\left(\beta{\cal M}+{\cal F}+{\cal
O}\right)\left(1+\sqrt{1+X^2}-\beta X\right)\\=\frac\beta2\left\{\left(1+\sqrt{1+X^2}\right),\left(2{\cal M}+\{{\cal O},X\}\right)\right\}+\bigl\{\sqrt{1+X^2}\left(1+\sqrt{1+X^2}\right),{\cal F}\bigr\}\\+\frac\beta2\left[X,[X,{\cal
M}]\right]-\frac12[X,[X,{\cal
F}]]\\+\frac\beta2\left\{\left(1+\sqrt{1+X^2}\right),[X,{\cal F}]\right\}-\frac\beta2\left\{X,[\sqrt{1+X^2},{\cal F}]\right\}. \end{array}$$
The transformed Hamiltonian is given by
\begin{equation}\begin{array}{c}
{\cal H}'=\beta\epsilon+ {\cal E}+\frac 14\left\{\frac{1}
{2\epsilon^2+\{\epsilon,{\cal M}\}},\left(\beta\left[{\cal O},[{\cal O},{\cal
M}]\right]-[{\cal O},[{\cal O},{\cal
F}]]\right)\right\}+{\cal O}',\\
{\cal O}'=\frac\beta4\left\{\frac{1}{\sqrt{1+X^2}},[X,{\cal
F}]\right\}-\frac\beta4\left\{\frac{X}{\sqrt{1+X^2}\left(1+\sqrt{1+X^2}\right)},[\sqrt{1+X^2},{\cal
F}]\right\},
\end{array} \label{MHamt}
\end{equation} where $\epsilon$ is defined by Eq. (\ref{epsln}). The last commutator can be
approximated as
$$[\sqrt{1+X^2},{\cal F}]\approx\left\{\frac{1}{2\sqrt{1+X^2}},[X^2,{\cal F}]\right\}.$$

Since the remaining odd term, ${\cal O}'$, is proportional to $\hbar$, it is comparatively small. This term can be eliminated thanks to a second transformation with the exponential operator
which is given by
\begin{equation}\begin{array}{c}
S'=-\frac{i\beta}{4}\left\{\frac{1}{\epsilon},{\cal O}'\right\}=-\frac{i}{16}\left\{\frac{1}{\epsilon},\left\{\frac{1}{\sqrt{1+X^2}},[X,{\cal F}]\right\}\right\}\\+\frac{i}{32}\left\{\frac{1}{\epsilon},\left\{\frac{X}{(1+X^2)(1+\sqrt{1+X^2})},[X^2,{\cal F}]\right\}\right\}. \end{array} \label{excondw}
\end{equation}

Additional even terms appearing after the second transformation can be neglected. Therefore, the obtained Hamiltonian is even and equal to 
\begin{equation}\begin{array}{c}
{\cal H}''={\cal H}'-{\cal O}'=\beta\epsilon+ {\cal E}+\frac 14\left\{\frac{1}
{2\epsilon^2+\{\epsilon,{\cal M}\}},\left(\beta\left[{\cal O},[{\cal O},{\cal
M}]\right]-[{\cal O},[{\cal O},{\cal
F}]]\right)\right\}.
\end{array} \label{MHamf} \end{equation}

\section{Correction of the Foldy-Wouthuysen Hamiltonian}

To calculate the relativistic FW Hamiltonian, we need to determine corrections to ${\cal H}''$. The Hamiltonian (\ref{MHamf}) does not satisfy the Eriksen condition. This statement is based
on the Baker-Campbell-Hausdorff (BCH) formula \cite{BakerCampbellHausdorff}, which defines the product of two exponential operators:
\begin{eqnarray}
\exp(A)\exp(B)=\exp{\biggl(A+B+\frac12[A,B]+\frac{1}{12}[A,[A,B]]-\frac{1}{12}[B,[A,B]]}\nonumber\\ -\frac{1}{24}\bigl[A,[B,[A,B]]\bigr]+{\rm higher~ order~
commutators}\biggr).\label{EriKorl}\end{eqnarray}
The product of two exponential operators can be calculated with any needed accuracy \cite{Dynkin}.

When $A=iS,~B=iS'$, and the  operators $S,S'$ are odd and Hermitian ($\beta$-pseudo-Hermitian for bosons), the operators $[A,B]$ and $\bigl[A,[B,[A,B]]\bigr]$ are even. Therefore, the resulting transformation operator $U_{res}=\exp{(iS')}\exp{(iS)}$
can be presented in the form $U_{res}=\exp{(i\mathfrak{R})}$, where the operator $\mathfrak{R}$ is not odd and does not satisfy the Eriksen condition (\ref{VveEfrt}).

Any subsequent exponential operator of the iterative FW transformation is of a smaller order of magnitude than the preceding one. Therefore, the BCH formula allows us to determine and to eliminate an error. Such a possibility has been first noticed by Eriksen and Kolsrud \cite{EK}.
If one can neglect $[S,[S,S']]$, $[S',[S,S']]$, and commutators of higher orders as compared with $[S,S']$, Eq. (\ref{EriKorl}) leads to the approximate relation
\begin{eqnarray}
\exp(iS')\exp(iS)=\exp{\left(\frac12[S,S']\right)}\exp{[i(S'+S)]}.\label{EriKorlnew}\end{eqnarray}
Since the operator $i(S'+S)$ is odd and the Hamiltonian ${\cal H}''$ is even (with a needed accuracy), the left multiplication of the FW transformation operator by the even operator
\begin{eqnarray}
U_{corr}=\exp{\left(-\frac12[S,S']\right)} \label{FWcorro}\end{eqnarray} does not add any odd terms to the Hamiltonian and
allows us to cancel the error of the iterative FW method in the leading order. For the stationary case, it has been shown in Ref. \cite{EK}.
The final FW Hamiltonian is equal to
\begin{equation}
{\cal H}_{FW}=U_{corr}\left({\cal H}''-i\hbar\frac{\partial}{\partial
t}\right)U_{corr}^{-1}+ i\hbar\frac{\partial}{\partial t}={\cal H}''-\Biggl[\frac12[S,S'],\left({\cal H}''-i\hbar\frac{\partial}{\partial
t}\right)\Biggr]. \label{corrrFW} \end{equation} In this equation, only the leading correction is taken into account.
The commutator of the exponential operators is equal to
\begin{equation}
[S,S']=-\frac{i\beta}{2}\left\{\arctan{X},S'\right\}. \label{commtwe}
\end{equation}

Equations (\ref{excondw}), (\ref{corrrFW}), and (\ref{commtwe}) clearly show that corrections to the Hamiltonian (\ref{MHamf}) are of the order of $(\hbar/S_0)^2$ and can be neglected in the considered case.
As a result, \begin{equation} {\cal H}_{FW}={\cal H}'',  \label{equiv}
\end{equation} and the relativistic FW Hamiltonian for the arbitrary-spin particle in external fields is given by Eq. (\ref{MHamf}).
%
This equation coincides with Eq. (\ref{eqfingn}), but the validity of the Hamiltonian is now rigorously proven.

The result obtained solves the problem of the FW
transformation for the relativistic particle of arbitrary spin in arbitrarily
strong external fields. Contrary to previously developed relativistic methods, the present method satisfies the Eriksen condition of the exact FW transformation, and therefore, its validity is appropriately proven. The derived FW Hamiltonian is expanded in powers of the Planck constant and contains terms proportional to the zero and first
powers. These terms are in accordance with the Eriksen condition and are determined exactly.
The agreement between the formulas for the FW Hamiltonians obtained in Ref. \cite{PRA} and in the present work substantiates previously obtained results for single relativistic particles in external fields. However, additional terms of the order of $(\hbar/S_0)^2$ which are also presented in the final equations for the FW Hamiltonian obtained in Refs. \cite{B,JMP,TMP,PRA,Goss} are not relevant. Certainly, all corrections of the order of $(\hbar/S_0)^2$ to the Hamiltonian (\ref{MHamf}) can be equal to zero. In the next section, we will consider an example of the relativistic FW transformation illustrating this possibility.

Unfortunately, none of the known relativistic iterative methods admit a derivation of corrected \emph{relativistic} FW Hamiltonians including terms of second and higher order in $\hbar$. In particular, the method stated in this section does not give an explicit relativistic expression for the anticommutator $\{\arctan{X},S'\}$. As a result, the correction to the FW Hamiltonian (\ref{MHamf}) can be obtained only as a series on powers of $\{(1/{\cal M}),{\cal O}\}$ and $\{(1/{\cal M}),{\cal F}\}$.

Thus, the present method of the FW transformation does not use an expansion in powers of $v/c$ and does not need a weak-field approximation. An expansion in powers of $\hbar/S_0$ means that this method is applicable when the de Broglie wavelength is much smaller than the characteristic distance \cite{PRA}:
\begin{equation}
\frac{\hbar}{p}\ll l. \label{cndition} \end{equation}
For example, this condition is always satisfied for particles or nuclei in accelerators and storage rings. In this case, quantum-mechanical effects can be very important (see, e.g., Ref. \cite{Papinietal}).
It seems to be particularly important that the FW Hamiltonian (\ref{MHamf}) can be applied in variational approaches. The FW Hamiltonians expanded in powers of $v/c$ and obtained in the framework of the weak field approximation are useful only for perturbation calculations. Nevertheless, the relative smallness of the de Broglie wavelength may lead to some restrictions. When the condition (\ref{cndition}) is not satisfied, omitted terms of the order of $(\hbar/S_0)^2,~(\hbar/S_0)^3,\dots$ may not be small.

\section{Relativistic spin-1 particle in a uniform magnetic field}\label{Example}

In this section, we use the system of units $c=1$ while $\hbar$ is included in all equations.

As an example of application of the present method of the  FW transformation, let us consider a relativistic pointlike spin-1 particle  in a uniform magnetic field. Such a particle
is described by the Proca equations \cite{Pr} with the added Corben-Schwinger term \cite{CS} characterizing the anomalous magnetic moment. 
The Sakata-Taketani transformation \cite{SaTa} allows us to obtain the Hamiltonian form of the initial equations. The sign of the direct product $\otimes$ will hereinafter be omitted. The general Hamiltonian in the
Sakata-Taketani representation has been derived in Ref. \cite{YB}.
For a spin-1 particle in the magnetic field $\bm B$, it can be presented in the form \cite{TMPFW}
\begin{equation}  \begin{array}{c} {\cal H}=\rho_3 \mathfrak{M}+{\cal E}+{\cal
O}, ~~~\mathfrak{M}=m+\frac{\bm\pi^2}{2m}-\frac{e\hbar}{m}\bm
S\cdot\bm B, \\
{\cal E}=-\rho_3\frac{e\hbar(g-2)}{2m}\bm S\cdot\bm B, \\
{\cal O}=i\rho_2\left[\frac{\bm\pi^2}{2m}-\frac{(\bm\pi\cdot \bm
S)^2}{m}+\frac{e\hbar(g-2)}{2m}\bm S\cdot\bm B\right],
\end{array} \label{eqXVI} \end{equation}
where $g-2$ defines the anomalous part of the magnetic moment $\mu'=e(g-2)\hbar s/(2mc)$, $\bm\pi=-i\nabla-e\bm A$ is the kinetic momentum, and $\bm S$ is the $3\times3$ spin matrix. We imply that spin-independent terms are multiplied by the $3\times3$ unit matrix $\mathfrak{I}$. The even operators $\rho_3 \mathfrak{M}$ and ${\cal E}$ and the odd operator ${\cal O}$ are commutative and anticommutative with the matrix $\beta$, respectively.
The wave function in the Sakata-Taketani representation is a six-component analog of a Dirac bispinor.

The presented separation of the even part of ${\cal H}$ into $\rho_3\mathfrak{M}$ and ${\cal E}$ is very convenient because $[{\cal I}\mathfrak{M},{\cal E}]=[{\cal I}\mathfrak{M},{\cal O}]=0$.
As a result, the initial Hamiltonian (\ref{eqXVI}) is similar to the Dirac Hamiltonian (\ref{eq3Dirac}), and the results obtained with the Eriksen method can be applied.

We may restrict ourselves to the consideration of a particle with an anomalous magnetic moment moving in the plane orthogonal to the field direction. When the magnetic field is uniform and $\bm B=B\bm e_z$, the operator $\pi_z=p_z$ commutes with the Hamiltonian (\ref{eqXVI}). The FW transformation of the operator $\pi_z$ does not change its form. Therefore, this operator also commutes with the FW Hamiltonian and has eigenvalues ${\cal P}_z={\rm const}$. Consequently, a consideration of the particular case ${\cal P}_z=0$ is quite reasonable \cite{TMPFW}. The general case of a nonzero vertical momentum can be reduced to this one by an appropriate Lorentz transformation.

The use of Eq. (\ref{MHamf}) allows us to obtain the following FW Hamiltonian \cite{TMPFW}:
\begin{equation}\begin{array}{c} {\cal H}_{FW}=
\rho_3\epsilon-\rho_3\frac{e\hbar(g-2)}{2m} \bm
S\cdot\bm B\\+\rho_3\frac{e^2\hbar^2(g-1)(g-2)}{16m^3}
\Biggl\{\frac{1}{\epsilon(\epsilon+m)},\biggl(B^2(\bm S\cdot\bm
\pi)^2\\-[\bm S\cdot(\bm \pi\times\bm B)]^2-e\hbar(g-1)B^2(\bm
S\cdot\bm B)\biggr)\Biggr\},\\ \epsilon=\sqrt{m^2+\bm
\pi^{2}-2e\hbar\bm{S}\cdot\bm B-\frac{e^2\hbar^2g(g-2)}{4m^2}(\bm
S\cdot\bm B)^2}.
\end{array}\label{eqprf}
\end{equation}

In this case, $S_0=\epsilon^2/(|e|B)\sim
m^2/(|e|B)$. Therefore, Eq. (\ref{eqprf}) contains even terms of the order of $(\hbar/S_0)^3$. However, the precision is not exceeded.
Since we can apply the results obtained with the Eriksen method, we can evaluate corrections introduced by the next terms in the expansion of the FW Hamiltonian in powers of $\hbar$ with Eqs. (\ref{eq12erk}) and (\ref{eq12A}).
The operators in these equations satisfy the relations
\begin{equation}      \begin{array}{c}
[{\cal O}^2,{\cal E}]=0,~~~[{\cal O},{\cal E}]=\rho_1\frac{
e^2\hbar^2(g-1)(g-2)}{2m^2}(\bm S\cdot\bm B)^2, \\ \left\{{\cal
O},[[{\cal O},{\cal E}],{\cal E}]\right\}=-\frac{
e^4\hbar^4(g-1)^2(g-2)^2}{2m^4}B^2(\bm S\cdot\bm B)^2, \\
\left([{\cal O},{\cal E}]\right)^2=\frac{
e^4\hbar^4(g-1)^2(g-2)^2}{4m^4}B^2(\bm S\cdot\bm B)^2.
\end{array}\label{eqrel} \end{equation}
In Eqs. (\ref{eq12erk}) and (\ref{eq12A}), among the terms of a nominal order of $(\hbar/S_0)^2$, only two, which are proportional to
$\left\{{\cal O},[[{\cal O},{\cal E}],{\cal E}]\right\}$ and
$\left([{\cal O},{\cal E}]\right)^2$, are nonzero, and they are in fact of the fourth order in $\hbar$ and $B$.
Because $S_0=\epsilon^2/(|e|B)\sim
m^2/(|e|B)$ in the considered case, the relativistic FW transformation allows determining the Hamiltonian for
spin-1 particles up to terms of the order of $(|e|\hbar B)^3/m^5$ inclusively. It is difficult to find another method which would ensure such high precision in this case.
Certainly, the important possibility to derive the FW Hamiltonian with an accuracy of the order of $(\hbar/S_0)^2$ and higher exists only in some special cases.

Let us obtain the energy spectrum with an accuracy of the order of $(\hbar/S_0)^2$ to demonstrate the usefulness of the present method of the relativistic FW transformation. In this case, we should take into account terms proportional to $B^2$ and to squared spin operators. These terms characterize the tensor electric and magnetic polarizabilities of a \emph{moving} particle. We will apply results obtained in Refs.  \cite{TMPFW,EPJC,PRDexact}.

To solve the problem, it is convenient to omit terms of the order of $(\hbar/S_0)^3$ and to represent the FW Hamiltonian in the form  \cite{PRDexact}
\begin{equation}\begin{array}{c} {\cal H}_{FW}={\cal H}_0+\rho_3\omega_0S_z -\rho_3\frac{e^2g(g-2)}{8m^2\epsilon'}
S_z^2B^2\\-\rho_3\frac{e^2(g-1)(g-2)}{16m^3}\left\{\frac{\epsilon'-m}{\epsilon'},
(S_{\bm\pi\times\bm B}^2-S_{\pi}^2)\right\}B^2,\\ {\cal H}_0=\rho_3\sqrt{m^2+\bm
\pi^{2}-2eS_zB},~~~ \epsilon'=\sqrt{m^2+\bm\pi^2}, ~~~ \omega_0=-\frac{e(g-2)}{2m}B,
\end{array}\label{wqpqf}
\end{equation} where the spin matrices define projections of the spin operator onto the corresponding directions.

It can be proven \cite{EPJC} that the operators $S_z,~S_{\pi}$, and $S_{\bm\pi\times\bm B}$ commute with ${\cal H}_0$. Therefore, the operator ${\cal H}_0$ commutates with the total FW Hamiltonian ${\cal H}_{FW}$.

For a particle with the normal magnetic moment ($g=2$), the operator $\bm\pi^2$ commutates with the FW Hamiltonian and its eigenvalues are defined by the Landau formula:
\begin{equation}\begin{array}{c} \bm\pi^2_n=|e|B(2n+1), ~~~ n=0,1,2,\dots
\end{array}\label{eigenpi}\end{equation} The eigenvalues of the operator ${\cal H}_0$ are given by \cite{TMPFW}
\begin{equation}      \begin{array}{c}
{\cal H}_0^{(n\lambda)}=\sqrt{m^2+(2n+1)|e|\hbar B-2\lambda e\hbar B}, ~~~
n=0,1,2,\dots,~~~ \lambda=+1,0,-1.
\end{array}\label{eq16e} \end{equation}

As a result, the eigenvalues of this operator are degenerate. When $n-es_z/|e|\ge1$, the degree of their degeneracy is equal to 3. The reason for the degeneracy is the commutation of ${\cal H}_0$ with all spin components ($S_z,~S_{\pi}$, and $S_{\bm\pi\times\bm B}$).
The energy levels (except for the lower level) of a Dirac particle with $g=2$ in a uniform magnetic field are twice degenerate.

In the considered case ($g\neq2$), the eigenvalues of the operator ${\cal H}_0$ are also defined by Eq. (\ref{eq16e}) but the operators $\bm\pi^2$ and $S_z$ do not commute with ${\cal H}_{FW}$ and do not have eigenvalues separately. Thus, the particle polarization in stationary states is defined by the remaining terms in the total FW Hamiltonian. It has been found in Ref. \cite{PRDexact} that
\begin{equation}\begin{array}{c} <\pm1|S_{z}|\pm1>=\pm Y,~~~<\pm1|S_{\bm \pi\times\bm B}^2|\pm1>=\frac{1\pm\mathfrak{B} Y}{2},\\
<\pm1|S_{\pi}^2|\pm1>=\frac{1\mp\mathfrak{B} Y}{2},~~~<\pm1|S_{z}^2|\pm1>=1,\\
<0|S_{z}|0>=0,~~~<0|S_{\bm \pi\times\bm B}^2|0>=<0|S_{\pi}^2|0>=1,\\
~~~<0|S_{z}^2|0>=0,~~~ Y=\frac{1}{\sqrt{1+\mathfrak{B}^2}}, ~~~ \mathfrak{B}=
\frac{e(g-1)(\epsilon'-m)}{4m^2\epsilon'}B.
\end{array}\label{eigenpp}\end{equation}
In any stationary state,
\begin{equation}\begin{array}{c} <S_{\bm \pi\times\bm B}>=<S_{\pi}>=0, \\  \left\langle\{ S_{\bm \pi\times\bm B},S_{\pi}\}\right\rangle=\left\langle\{S_{\bm \pi\times\bm B},S_{z}\}\right\rangle=\left\langle\{S_{\pi},S_{z}\}\right\rangle=0.
\end{array}\label{additio}\end{equation}

While the spin projections in two stationary states are not integer, we retain the traditional designation of spin states. With the required precision, $\epsilon'$ is determined with Eq. (\ref{eq16e}).

Equations (\ref{wqpqf}) and (\ref{eigenpp}) define the eigenvalues of ${\cal H}_{FW}$ and therefore the energy spectrum:
\begin{equation}\begin{array}{c} {\cal H}_{FW}\Psi_\lambda=E_\lambda\Psi_\lambda~~~(\lambda=+1,0,-1),\\
E_{\pm1}={\cal H}_0^{(n,\pm1)}\pm\omega_0\sqrt{1+\mathfrak{B}^2}-\frac{e^2g(g-2)}{8m^2\epsilon'}B^2,~~~E_{0}={\cal H}_0^{(n0)}. 
\end{array}\label{eigenvv}\end{equation}

To analyze the energy spectrum, the degeneracy of the energy levels of ${\cal H}_0^{(n\lambda)}$ should be taken into account. It is convenient to take $n|e|-\lambda e=const~(\lambda=+1,0,-1)$ and therefore to substitute the values of ${\cal H}_0^{(n\lambda)}$ independent of $n$ and $\lambda$ into the expressions for $E_\lambda$.

We can conclude that the relativistic FW transformation has allowed us to obtain the high-precision equation for the energy spectrum of pointlike spin-1 particles (W$^\pm$ bosons) in a uniform magnetic field. Equation (\ref{eigenvv}) clearly demonstrates a rather nontrivial form of the contribution of the tensor polarizabilities to the energy spectrum.  This example shows the great possibilities opened by the relativistic FW transformation for finding important physical properties of particles in external fields.

\section{Summary}

We have presented a new method of the FW transformation that is useful for relativistic particles with any spin in arbitrarily strong external fields. Contrary to previously developed relativistic methods \cite{B,JMP,TMP,BliokhK,PRA,Goss}, it is rigorous and satisfies the Eriksen condition of the exact FW transformation. This method is general because it is applicable to relativistic particles with any spin in arbitrarily strong external fields.
The final equation (\ref{MHamf}) is useful when the de Broglie wavelength is much smaller than
the characteristic distance.
The relativistic FW Hamiltonian is expanded in powers of the Planck constant. In this Hamiltonian, terms proportional to the zero and first
powers of the Planck constant satisfy the Eriksen condition and are determined exactly, while lower-order terms are not specified. Let us mention that the Eriksen condition was not used by Eriksen or in other works even for a relativistic or ultrarelativistic Dirac particle, let alone for a particle with another spin. We also note that the relativistic method of
the FW transformation for a Dirac particle in external fields
developed in Ref. \cite{Morpurgo} satisfies the Eriksen condition.

The derived equation (\ref{MHamf}) agrees with the corresponding formula for the FW Hamiltonian previously deduced in Ref. \cite{PRA} with the iterative relativistic method and rigorously proves its validity. This substantiates previously obtained results for single relativistic particles in external fields.
The present approach gives a convergent expression for any particle energy, while the applicability of methods expanding the FW Hamiltonian into a series of relativistic corrections \cite{E,dVFor,QuantChem} is restricted by the condition $\bm p^2/(mc)^2<1$.

If a priori information about commutation relations stated in Sec. \ref{Methods} is right, corrections to the Hamiltonian (\ref{MHamf}) are proportional to the second and higher powers of the Planck constant. The example presented in Sec. \ref{Example} shows that these corrections can be less is some specific cases. This example demonstrates a usefulness of the relativistic FW transformation for a high-precision description of an arbitrary-spin particle in external fields.

\section*{Acknowledgements}

The author is indebted to G. Morpurgo for interest in the
present work and for his valuable comment. This work was supported in part by the Belarusian Republican Foundation for
Fundamental Research (Grant No. $\Phi$14D-007) and by the Heisenberg-Landau program of the German Ministry for Science and Technology (Bundesministerium f\"{u}r Bildung und Forschung).


\end{document}